\documentclass[epj]{webofc}
\usepackage[utf8]{inputenc} 
\usepackage[varg]{txfonts}   
\usepackage{booktabs}
\usepackage{xcolor}
\definecolor{darkred}{rgb}{0.4,0.0,0.0}
\definecolor{darkgreen}{rgb}{0.0,0.4,0.0}
\definecolor{darkblue}{rgb}{0.0,0.0,0.4}
\usepackage[bookmarks,linktocpage,colorlinks,
    linkcolor = darkred,
    urlcolor  = darkblue,
    citecolor = darkgreen]{hyperref}
\usepackage{subfigure}
\wocname{EPJ Web of Conferences}
\woctitle{Lattice2017}
\newcommand{\dd}{\mathrm{d}}
\usepackage{multirow}
\usepackage{bbm}
\usepackage[justification=centering]{caption}
\begin{document}
%
\selectlanguage{english}
\title{%
Thermal modifications of charmonia and bottomonia from spatial correlation functions
}
\author{%
\firstname{Heng-Tong} \lastname{Ding}\inst{1} \and                                   
\firstname{Olaf} \lastname{Kaczmarek}\inst{1,2} \and
\firstname{Anna-lena} \lastname{Kruse}\inst{2} \and
\firstname{Swagato} \lastname{Mukherjee}\inst{3} \and
\firstname{Hiroshi} \lastname{Ohno}\inst{3,4} \and
\firstname{Hauke} \lastname{Sandmeyer}\inst{2} \and
\firstname{Hai-Tao}  \lastname{Shu}\inst{1}\fnsep\thanks{Speaker, \email{haitaoshu@mails.ccnu.edu.cn}}
}
\institute{%
Central China Normal University, Wuhan 430079, China
\and
Fakult\"at f\"ur Physik, Universit\"at Bielefeld, 33615 Bielefeld, Germany
\and
Physics Department, Brookhaven National Laboratory, Upton, NY 11973, USA
\and
Center for Computational Sciences, University of Tsukuba, Tsukuba, Ibaraki 305-8577, Japan
}
\abstract{%
  We present our study on the thermal modifications of charmonia and bottomonia from spatial correlation functions at zero and nonzero momenta in quenched QCD. To accommodate the heavy quarks on the lattice we performed simulations on very fine lattices at a fixed beta value corresponding to a lattice spacing $a^{-1}=22.8$ GeV on $192^3\times32$, $192^3\times48$, $192^3\times56$, $192^3\times64$ and $192^3\times96$ lattices using clover-improved Wilson fermions. These lattices correspond to temperatures of $2.25 T_c$, $1.50 T_c$, $1.25 T_c$, $1.10 T_c$ and $0.75 T_c$. To increase the signal to noise ratio in the axial-vector and scalar channels we used multi-sources for the measurement of spatial correlation functions. By investigating on the differences between spatial and temporal correlators as well as the temperature dependence of screening masses we will discuss the thermal effects in different channels of quarkonium states. Besides this the dispersion relation of the screening mass at different momenta is also discussed.
}
\maketitle
\section{Introduction}\label{intro}

In heavy-ion physics much effort is put into investigating the properties of quark-gluon plasma (QGP) formed after the collision. Heavy quarkonium states, which serve as a good QGP thermometer \cite{Matsui:1986dk}, are powerful probes to study the properties of the hot medium. In recent years enormous progress has been made experimentally in exploring the fate of heavy quarkonia embedded in the thermal medium. For instance, the Pb-Pb collisions carried out at CMS with $\sqrt{s_{NN}}=2.76$ TeV show a significant sequential suppression for bottomonium compared with that in the p-p collision \cite{Chatrchyan:2012lxa}. This phenomenon is believed to signal the presence of QGP. To have a better understanding of such a phenomenon, a detailed theoretical study on the behavior of the quarkonium states and their dissociation temperatures is needed. Currently there are two different strategies dealing with this problem \cite{Ding:2015ona}. One strategy is to study the spectral functions (SPFs) either extracted from correlation functions calculated on the lattice or obtained by solving a non-relativistic Schr$\ddot{o}$dinger equation with $q\bar{q}$ potential in lattice QCD simulations. However, studies using internal energy \cite{0954-3899-32-3-R01} and free energy \cite{Mocsy:2007yj} as the potential give different dissociation temperatures for quarkonia. Other studies suggest that the potential could be complex \cite{Laine:2006ns,Brambilla:2008cx,Beraudo:2007ky}, but the real part and imaginary part are still not well determined (see  \cite{Burnier:2014ssa}). SPFs can be extracted from the temporal correlation functions in Euclidean space-time calculated on the lattice by inversion methods \cite{Asakawa:2000tr,Burnier:2013nla,Brandt:2015aqk,Ohno:2016ggs,Shu:2015tva}. But due to the fact that the extraction is ill-posed one needs extra prior information about the SPF, which can introduce some uncertainties.

The other strategy is to study the temperature dependence of the screening mass. Quarkonia living in the hot medium are subject to the color screening effect and their thermal modifications can be reflected by the so-called screening mass. By examining the screening masses at different temperatures, the study of thermal effects on the quarkonia is available. One should note another important fact that the quarkonia produced in heavy-ion collisions are not at rest with respect to the medium usually. For example $J/\psi$ does not flow with the medium until $p_{\perp}\sim 8$ GeV/c \cite{Tang:2011kr}. Thus, it is necessary to carry out the study on the momentum dependence as well. Studies on this using effective theories or AdS/CFT can be found in \cite{Escobedo:2011ie,Liu:2006nn}. There are also lattice studies on non-zero momentum effects, see \cite{Aarts:2012ka,Ikeda:2016czj,Oktay:2010tf,Ding:2012pt}. In this paper, the results based on lattice QCD calculations on large quenched lattices are presented.

\section{Euclidean Spatial Correlator and Screening Mass}\label{screening mass}

Integrating the current-current correlation functions over $x$, $y$ and $\tau$ directions gives the spatial correlation function $G_H(z,\mathbf{p_{\perp}},\omega_n)$ which can be calculated directly on the lattice
\begin{align}
    \label{spatial_corr}
    G_H(z,\mathbf{p_{\perp}},\omega_n)= \sum_{x,y,\tau}\exp(-i \mathbf{\tilde{p}}\cdot\mathbf{\tilde{x}})\big{\langle} J_H(0,\mathbf{0})J^{\dag}_H(\tau,\mathbf{x})\big{\rangle}\ ,
\end{align}
where $\mathbf{p_{\perp}}=(p_x,p_y)$, $\omega_n=2n\pi T$ and $\mathbf{\tilde{p}}=(\mathbf{p_{\perp}},\omega_n)$. Here the current is defined as
\begin{align}
    \label{current}
    J_H(\tau,\mathbf{x})=\bar \psi(\tau,\mathbf{x})\Gamma_H\psi(\tau,\mathbf{x}) ,
\end{align}
where $\Gamma_{H}=\gamma_5, \gamma_{\mu}, \mathbbm{1}, \gamma_5\gamma_{\mu}$ corresponds to pseudo-scalar (P), vector (V), scalar (S) and axial-vector (A) channel, respectively. As the temporal dimension in lattice QCD is bound to $1/T$, it is reasonable to look at the spatial direction. The relation between the spatial correlation function and the spectral function is 
\begin{align}
    \label{spatial_corr_spf}
    G_H(z,\mathbf{p_{\perp}},\omega_n)=\int\limits_{-\infty}^{\infty}\frac{\dd p_z}{2\pi}\exp(ip_zz) \int\limits_0^{\infty}\frac{\dd \omega}{\pi} \rho_H(\omega,\mathbf{p},T)\frac{\omega}{\omega^2+\omega_{n}^2}\ ,
\end{align}
where $\mathbf{p}=(p_x,p_y,p_z)$. The spectral function appeared here is the same as the one which can be extracted from the temporal correlation function, but its extraction from the spatial correlation function is almost impossible because of the complexity. However, with little effort one could find that at large distance $z\rightarrow \infty$, the spatial correlation shows an exponential decay behavior characterized by an energy $E_{scr}$
\begin{align}
    \label{spatial_corr_decay}
    G_H(z,\mathbf{p_{\perp}},\omega_n)\sim \exp(-zE_{scr})\ ,
\end{align}
where $E_{scr}$ is given by
\begin{align}
    \label{absorb}
    E_{scr}^2=A(T)\mathbf{\tilde{p}}^2+M_{scr}^2(T)
\end{align}
which is inspired from the finite-temperature field theory calculation result
\begin{align}
    \label{FTFT}
    E_{scr}^2=\mathbf{\tilde{p}}^2+M_{scr}^2(T)+\Pi(\mathbf{\tilde{p}},T)
\end{align}
by absorbing the self-energy term $\Pi(\mathbf{\tilde{p}},T)$ into a $T$-dependent coefficient $A(T)$. At vanishing momentum limit $\mathbf{\tilde{p}}=0$, $E_{scr}$ is nothing but the screening mass $M_{scr}(T)$ and at zero temperature $T=0$, it is the same as the pole mass. In the limit of $T\rightarrow \infty$, the heavy quarkonia are dissolved and can be considered as a pair of free quark and its anti-quark. Then, the screening masses in this limit can be calculated in the non-interacting limit \cite{Florkowski:1993bq}
\begin{align}
    \label{screening_free}
    M_{scr}^{free}=2\sqrt{(\pi T)^2+m^2_q}\ .
\end{align}

The exponential decay behavior mentioned above becomes cosh-dependence on $z$ if a periodic boundary condition is adopted. In lattice QCD, $E_{scr}$ can be obtained by fitting to the spatial correlation according to Eq. (\ref{spatial_corr_decay}). The details of this procedure will be discussed in Sec.\ref{fit}. After determining $E_{scr}$ in different channels at different temperatures and momenta, the in-medium dispersion relation can be studied.

\section{Lattice Setup}\label{setup}

To calculate the spatial correlation functions of charmonia and bottomonia, we performed simulations on isotropic large quenched lattices with quite small lattice spacing $a^{-1}=22.8$ GeV. The lattice spacing $a$ is determined by using Sommer parameter $r_0$ \cite{Sommer:1993ce}. The spatial extent $N_{\sigma}$ is set to 192, which is large enough to make the fit of screening mass reliable. Temporal extent is widely ranged corresponding to temperatures from 0.75$T_c$ to 2.25$T_c$. We used non-perturbatively Clover-improved Wilson fermions \cite{Sheikholeslami:1985ij} and quark masses are tuned to reproduce nearly physical $J/\psi$ mass and $\Upsilon$ mass (see Fig. \ref{dispersion_relation}). To study the momentum dependence of the dispersion relation, the momentum is set in a range from 0$\sim$3.17 GeV. To increase the signal-to-noise ratio at zero momentum, we increase the statistics by using multiple sources (with mostly 5 source points for each gauge configuration), which reduces the relative error $\delta G/\bar{G}$ by approximately a factor of 2 compared with single source. At non-zero momenta, we use a single source and $\delta G/\bar{G}\sim 1.5\%$ at the middle point in the vector channel. The details of our simulations are summarized in Table.\ref{xxxxx}.

\begin{table}[thb]
\captionsetup{singlelinecheck = false, justification=justified}
  \small
  \centering
\begin{tabular}{|c|c|c|c|c|c|c|}
\hline
$\beta$ \ & $a^{-1}$ \ & $\kappa$ \ & $N_{\sigma}$ \ & $N_{\tau}$ \ & $T/T_c$ \ & $\#$conf(single-source/multi-source)\\ 
\multirow{4}*{7.793} & \multirow{4}*{22.8 GeV}& \multirow{2}*{0.13221($c\bar{c}$)} & \multirow{4}*{192} & 96 & 0.75 & 218/224 \\
                                                     ~ & & & & 64 & 1.10 & 248/291 \\
                                                     ~ & & & & 56 & 1.20 & 190/290 \\
                                                             ~ & & 0.12798($b\bar{b}$) & & 48 & 1.50 & 210/339 \\
                                                             ~ & & & & 32 & 2.25 & 235/235 \\
        \hline
    \end{tabular}
 \caption{Lattice spacing, quark mass, spatial extent, temperatures and statistics of configurations performed in this lattice simulations.}
  \label{xxxxx}
\end{table}

\section{Results}\label{results}

\subsection{Ratios of Correlators}\label{ratios}

Before showing the results for screening mass, let us look at the ratio of correlators which gives information about how the quarkonium state behaves at high temperature compared with that at temperature below $T_c$ in the medium. The main contribution to a correlation function comes from the ground state. Any temperature dependent change of the quarkonium state will be reflected on the variation of the correlation functions. Given two different temperatures, we can check whether the ratio of the correlators at these two temperatures is decreasing or increasing in $z$-direction. By making use of Eq. (\ref{spatial_corr_decay}), one is able to know at which temperature the screening mass is larger. Since the screening mass characterizes the long distance behavior we only focus on the large distance. As a simple example we compare the vector and axial-vector correlation functions for $b\bar{b}$ at zero momentum. For a better illustration we only show the ratio of correlators at temperature $T>T_c$ to that at $T=0.75T_c$.
\begin{figure}[htb]
\captionsetup{singlelinecheck = false, justification=justified}
   \centering
   {\includegraphics[width=0.475\textwidth,height=5.3cm]{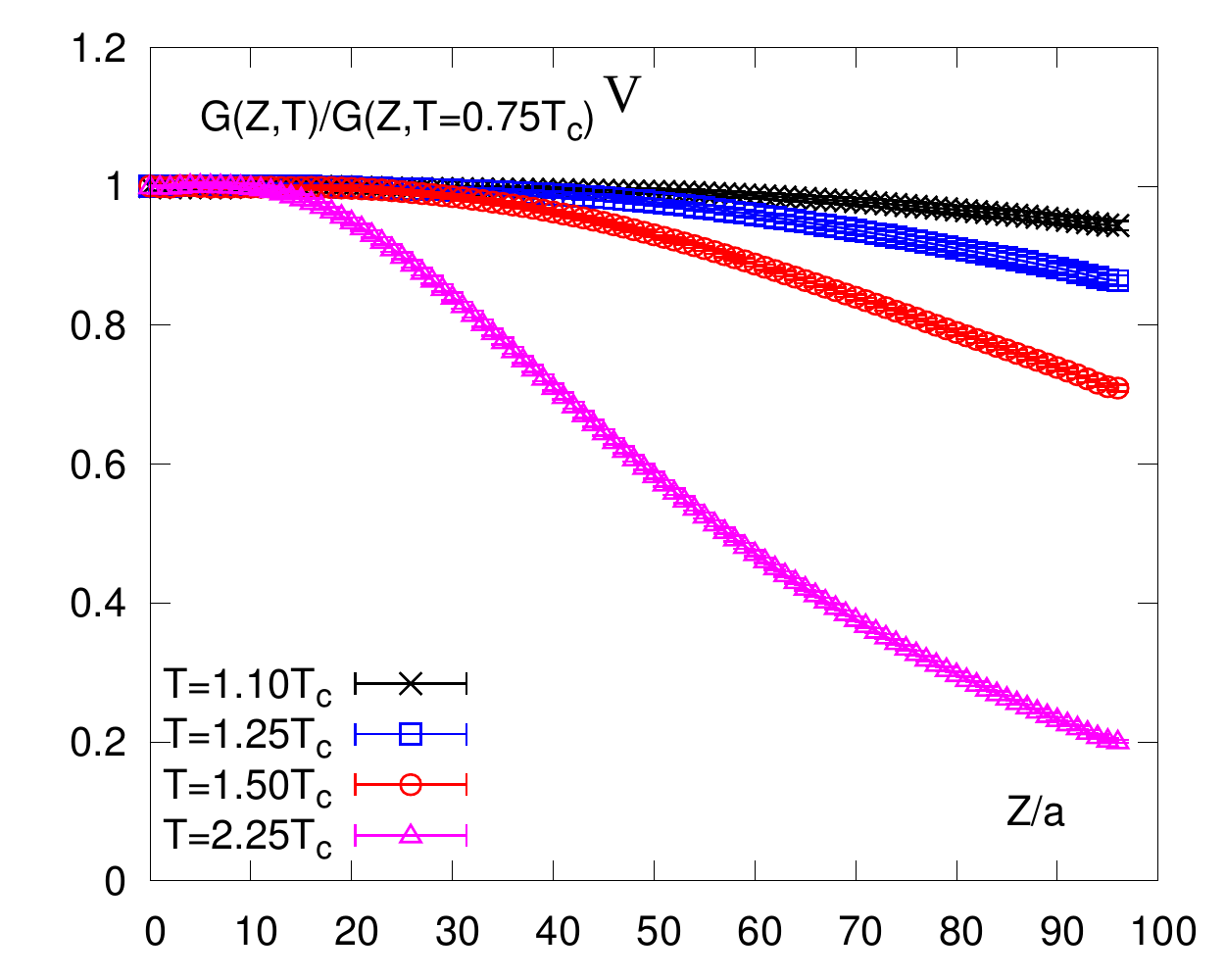}}\hfill
   {\includegraphics[width=0.475\textwidth,height=5.3cm]{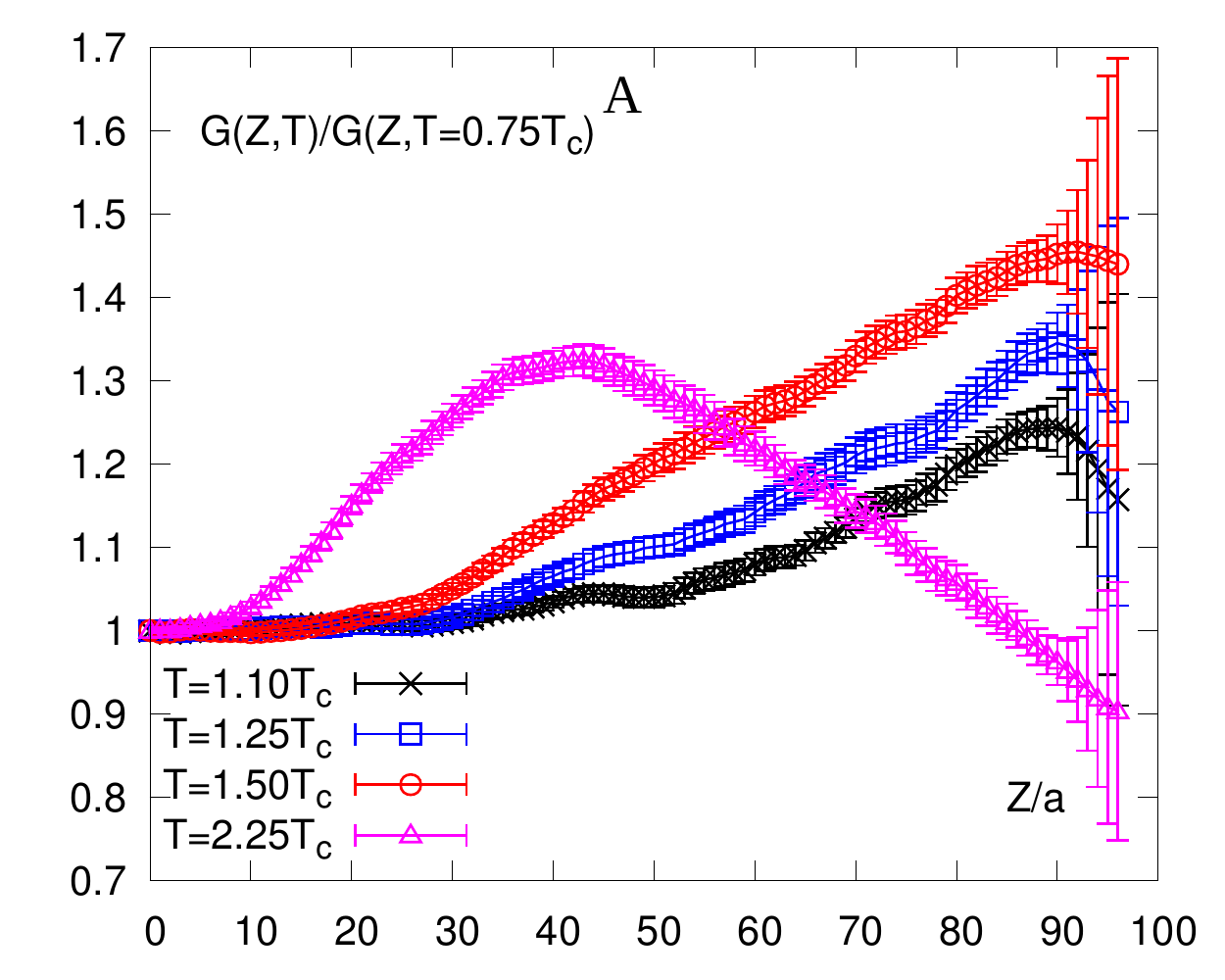}}
  
   \caption{The ratios of bottomonia correlation functions at different temperatures above $T_c$ to that at temperature $T=0.75T_c$ in the vector channel (\textit{Left}) and axial-vector channel (\textit{Right}).}
   \label{correlator_ratio_bb}
\end{figure}

First let us look at the vector channel shown in the left panel of Fig. \ref{correlator_ratio_bb}. We can see that the ratios are smaller than unity at all temperatures and decrease faster at higher temperature as a function of $z$. From the discussion above we easily know that the screening mass becomes larger as temperature increases. As for the axial-vector channel shown in the right panel, we find that the screening mass decreases at temperatures $1.10T_c$, $1.25T_c$ and $1.50T_c$. While at 2.25Tc the ratio has a non-monotonic behavior as a function of z which may indicate the largest thermal modification to the bound states.

Though not shown here, in our analysis we have found that the vector and axial-vector channel of $c\bar c$ behave similar to $b\bar b$ in the vector channel but the deviation of the ratios from unity is much larger. This indicates that charmonia suffer from more thermal modification than bottomonia in the medium.

\subsection{Fitting the Correlators}\label{fit}
In this section we consider how to extract the energy $E_{scr}$ from the spatial correlation function. To separate the contribution from different states we adopt a two-state fit ansatz
\begin{align}
\label{eq_two_state_fit}
G(n_{\sigma})=A_1\cosh[E^{2-state}_{scr1}(n_\sigma-N_{\sigma}/2)]+A_2\cosh[E^{2-state}_{scr2}(n_\sigma-N_{\sigma}/2)]\ .
\end{align}
Here $E_{scr}$ denotes the energy to be extracted in lattice unit and $n_{\sigma}=Z/a$. With this ansatz $E_{scr}$ can be obtained by performing a correlated $\chi^2$-fitting in the range [$n_{min},n_{max}$]. $n_{max}$ is determined by an empirical formula which gives optimal fit results. $n_{min}$ is varied successively from a small number to $n_{max}-4$ (we have 4 fit parameters) to generate different fit windows. By checking the $\chi^2/d.o.f$ in different windows, the best fit interval can be found when $\chi^2/d.o.f\sim 1$ and a clear plateau in $E^{2-state}_{scr2}$ can be seen in such an interval with a proper initial guess for fit parameters. Then the final results are obtained by averaging the data points in this interval.

\begin{figure}[htb] 
  \centering
  \sidecaption
  \includegraphics[width=0.55\textwidth,height=4.5cm]{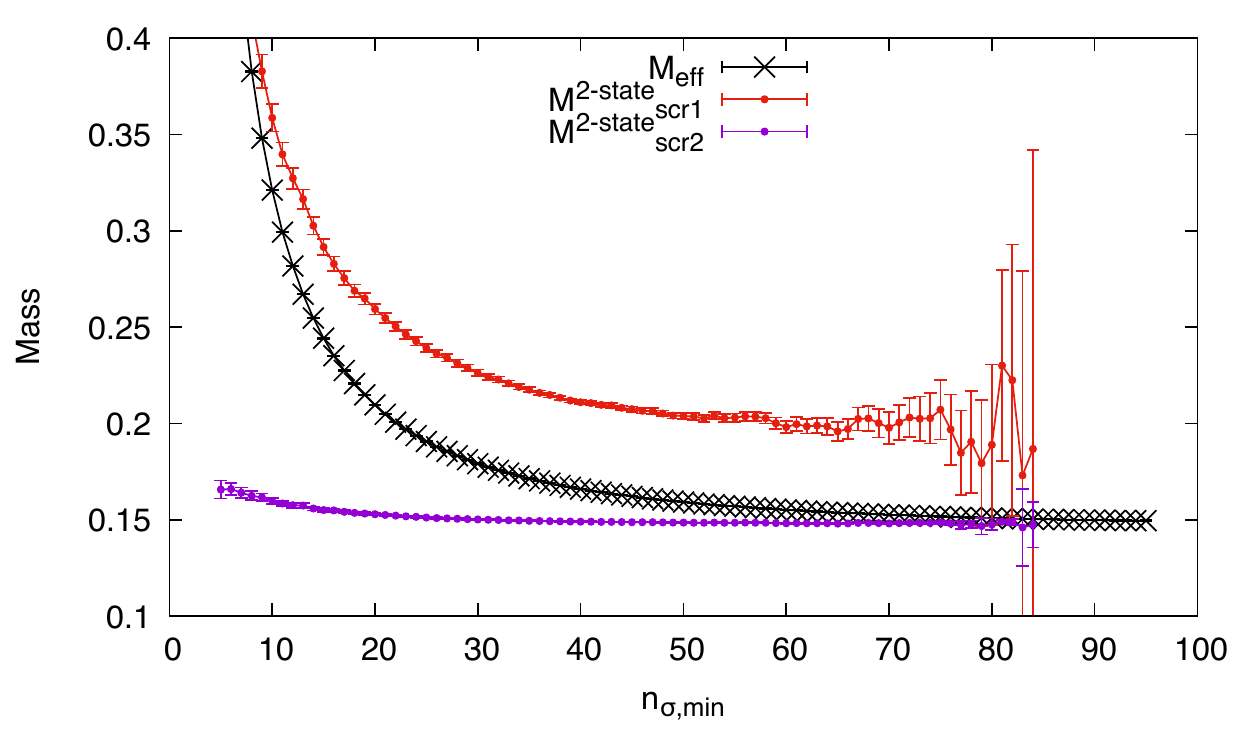}
  \caption{The comparison of $J/\psi$’s screening masses obtained in a correlated two-state fit with its effective mass.}
  \label{one_two_state}
\end{figure}

As an example, we applied the correlated two-state fit into the $J/\psi$ spatial correlation functions at $N_{\tau}=96$ with zero momentum, where the energy $E_{scr}$ becomes the screening mass $M_{scr}$. Fig. \ref{one_two_state} shows the screening masses and also the effective mass $M_{eff}$ (equal to $E_{eff}$ at zero momentum) which is obtained by solving the following equation
\begin{align}
\label{eq_mcosh}
\frac{G(n_{\sigma})}{G(n_{\sigma}+1)}=\frac{\cosh[E_{eff}(n_\sigma-N_{\sigma}/2)]}{\cosh[E_{eff}(n_\sigma-N_{\sigma}/2+1)]}\ .
\end{align}
We see that a clear plateau appears in $M^{2-state}_{scr2}$ and is close to $M_{eff}$ at the largest distance. 

\subsection{Screening Mass}\label{screening mass}

\begin{figure}[htb]
\captionsetup{singlelinecheck = false, justification=justified}
   \centering
             {\includegraphics[width=0.475\textwidth,height=5.5cm]{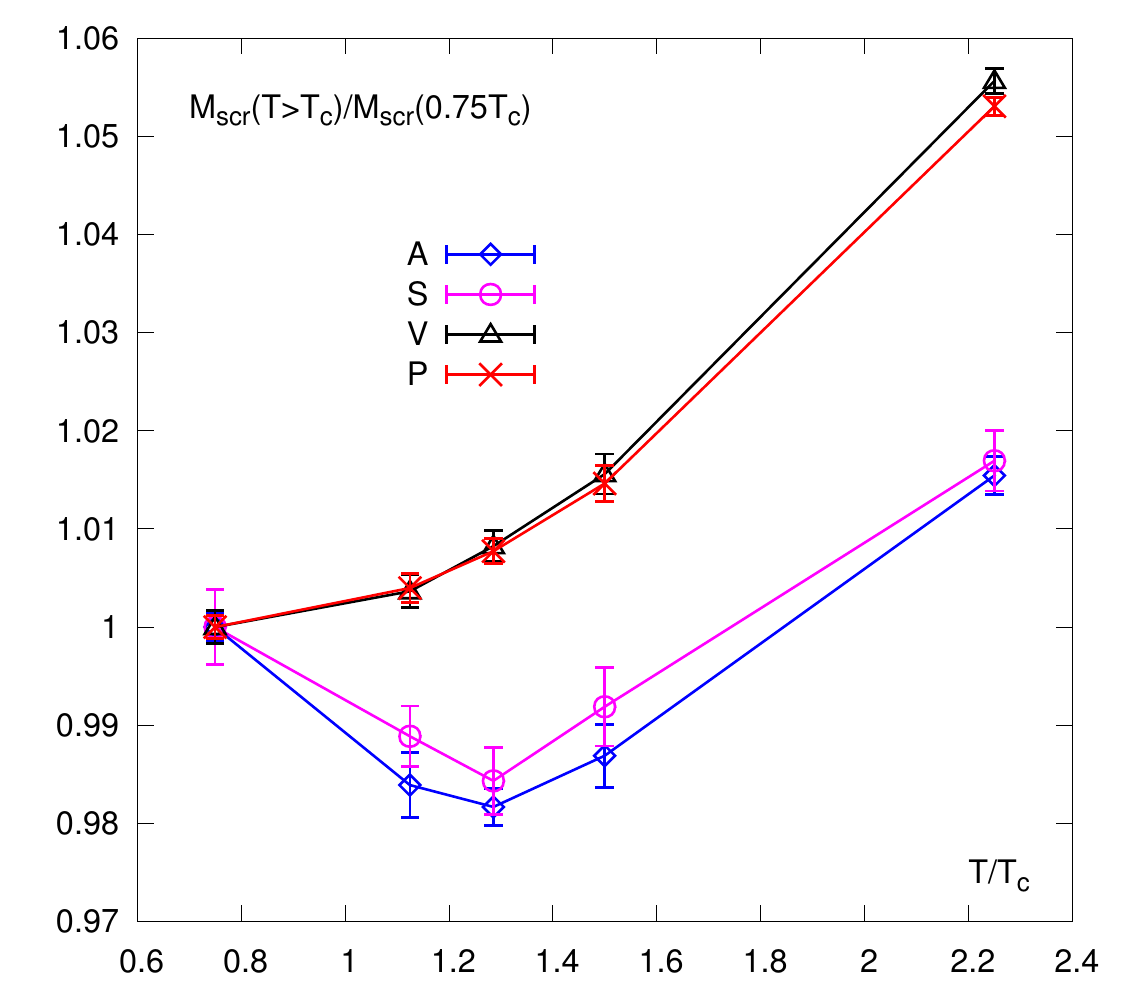}}\hfill
             {\includegraphics[width=0.475\textwidth,height=5.5cm]{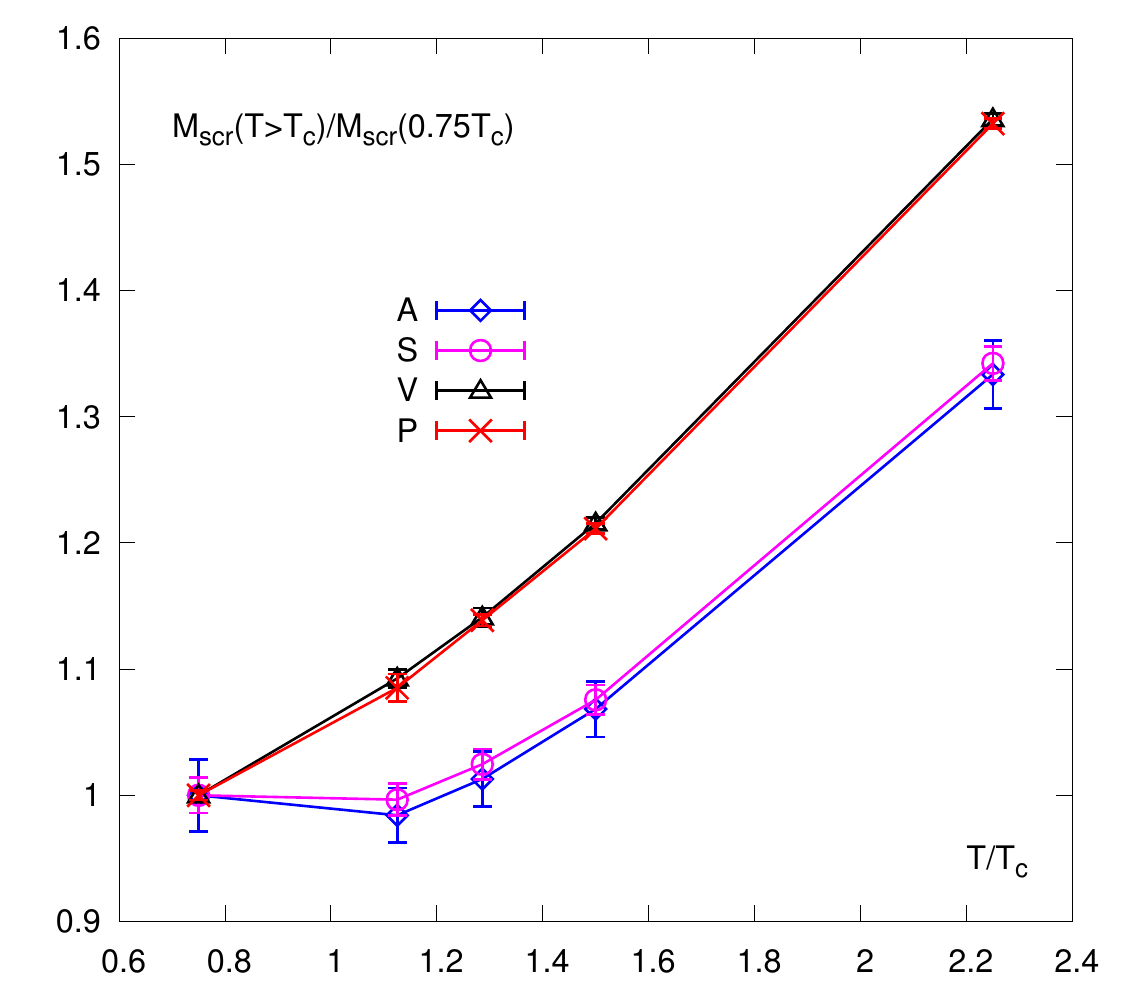}}
  
   \caption{Ratios of screening masses at different temperatures to that at temperature $T=0.75T_c$ in different channels. The label “A”, “S”, “P”, “V” denote axial-vector, scalar, pseudo-scalar, vector channel, respectively. \textit{Left:} Results of bottomonia. \textit{Right:} Results of charmonia.}
   \label{MT1_over_MT0}
\end{figure}

In this section we discuss the temperature dependence of the screening mass. Instead of showing the screening mass itself, we show the ratios of screening masses at different temperatures to that at temperature $T=0.75T_c$, which gives better demonstration of how the screening mass changes with temperature.

The left panel of Fig. \ref{MT1_over_MT0} shows the results for bottomonia. We can see that the ratios in the S and P channels are similar to those in the A and V channels, respectively. The screening masses of S-wave states increase monotonically by $\sim 5.6\%$ at $2.25T_c$ while for P-wave states they drop first and then go up. The right panel shows that a similar behavior is observed for charmonia as bottomonia. But the screening masses of S-wave states increase by $\sim 54\%$ at $2.25T_c$, much larger than that in the case of bottomonia. Similarly, for the P-wave states, the screening masses of charmonia also increase more than bottomonia at $2.25T_c$.  We conclude that charmonia suffer from more thermal effects than bottomonia in the medium. The discussions above verified our statements given in Sec. \ref{ratios} from our analysis of the correlation functions.

\begin{figure}[htb]
\captionsetup{singlelinecheck = false, justification=justified}
   \centering
             {\includegraphics[width=0.475\textwidth,height=5.5cm]{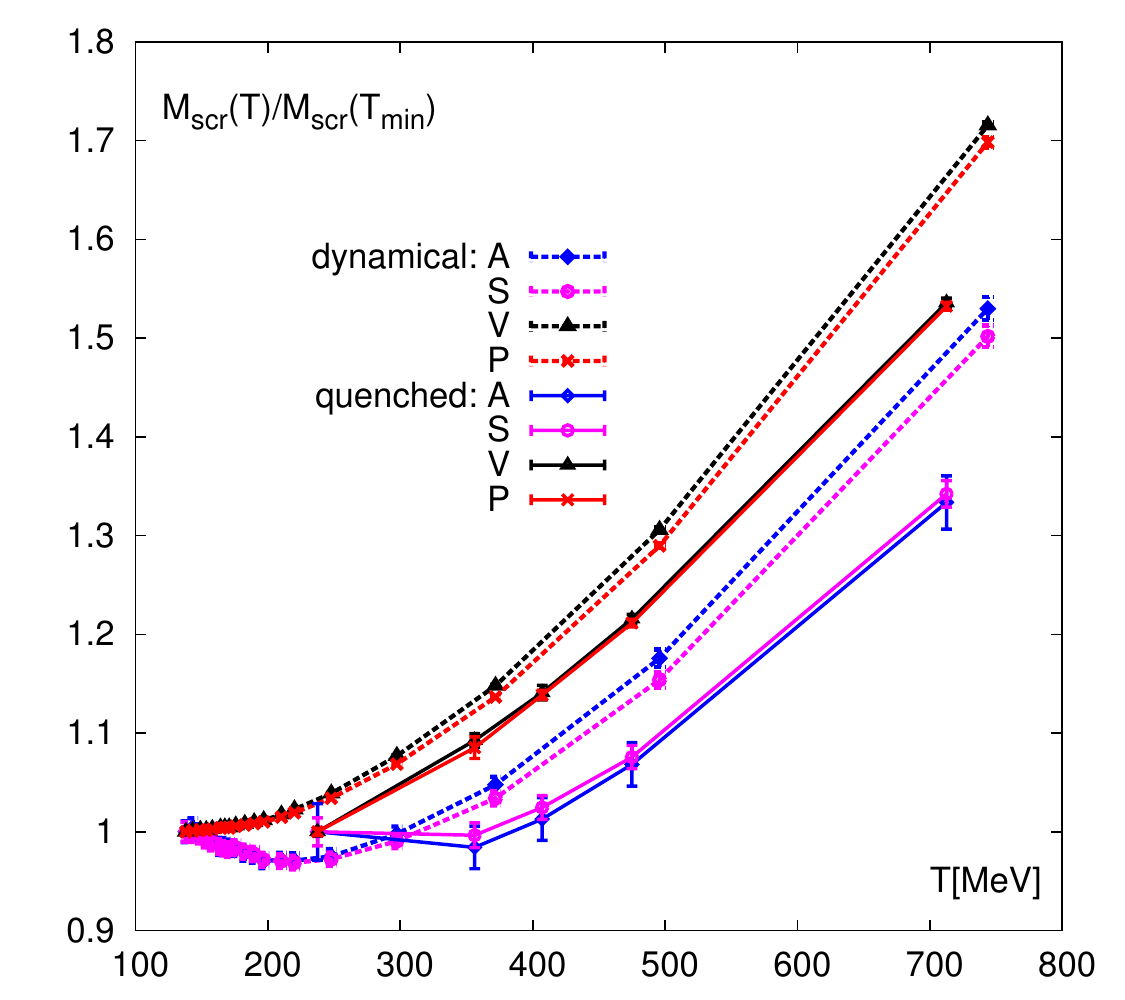}}\hfill
{\includegraphics[width=0.475\textwidth,height=5.5cm]{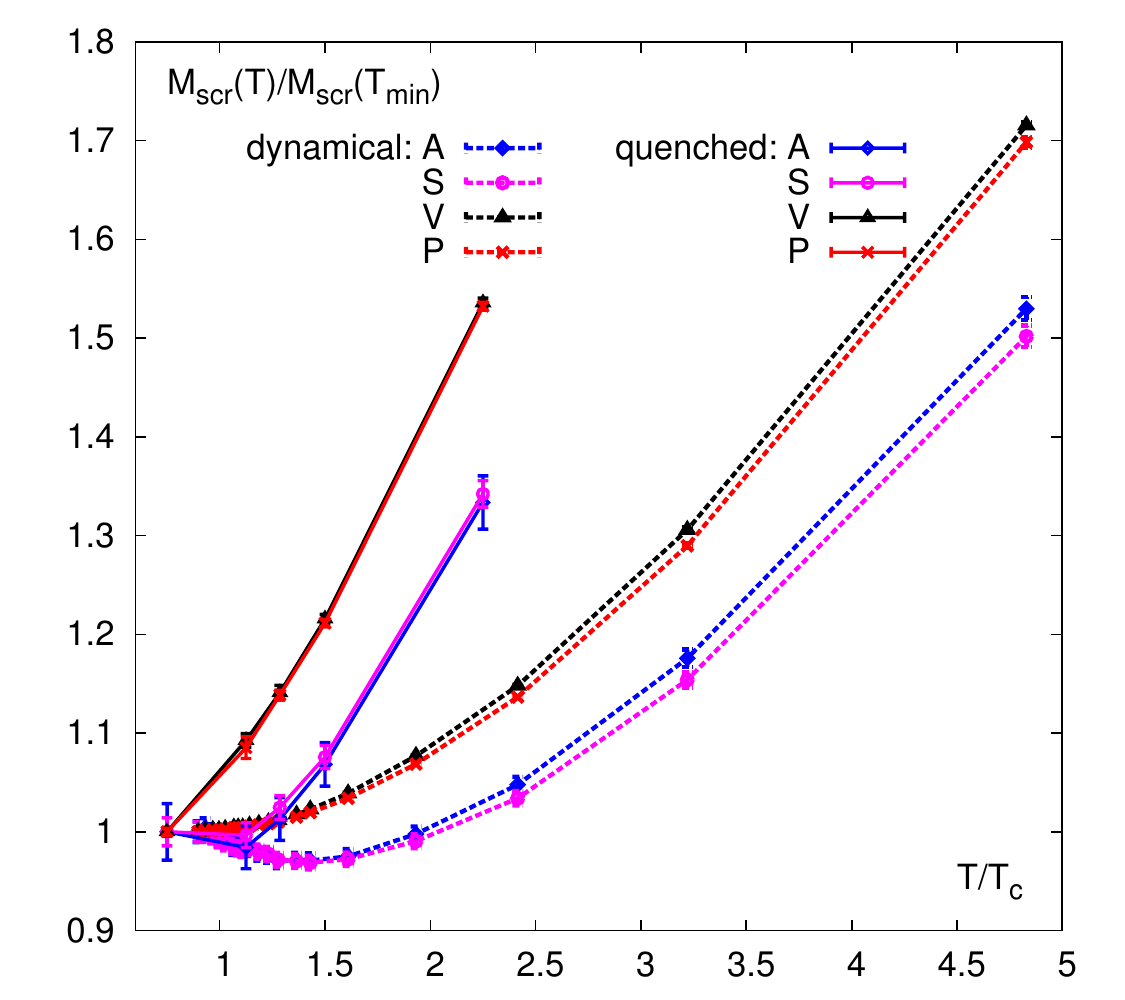}}\hfill
  
   \caption{Comparison of the screening mass of charmonia from this study with that from 2+1 flavor HISQ simulations \cite{Bazavov:2014cta}. The data points connected with solid lines are the results from our quenched simulations while the dashed lines are for the HISQ simulations. The left panel and right panel are the same except that the X-axis in the left panel is temperature $T$ while in the right panel it is $T/T_c$. $T_c$ is 313.7 MeV \cite{Francis:2015lha} for quenched and 154 MeV \cite{Bazavov:2011nk} for 2+1 flavor QCD. $T_{min}$ in the Y-axis means the lowest temperature in the simulations.}
   \label{quench_dynamic}
\end{figure}

We also compare the results of our quenched calculations with those obtained from $N_f=2+1$ lattice QCD calculations using the HISQ action \cite{Bazavov:2014cta}. As shown in Fig. \ref{quench_dynamic}, the screening masses of charmonia for both quenched and the $N_f= 2 + 1$ cases show the similar $T$-dependence: (I) for the S-wave states, they increase monotonically in temperature; (II) for the P-wave states, they decrease first and then go up as temperature increases. The difference is that in our quenched calculations, the screening masses for P-wave states have a dip at $1.10T_c$ for P-waves while in dynamic QCD the dip is at $1.43T_c$.

\subsection{Dispersion Relation}\label{dispersion relation}
\begin{figure}[tp]
\captionsetup{singlelinecheck = false, justification=justified}
   \centering
             {\includegraphics[width=0.475\textwidth,height=5.3cm]{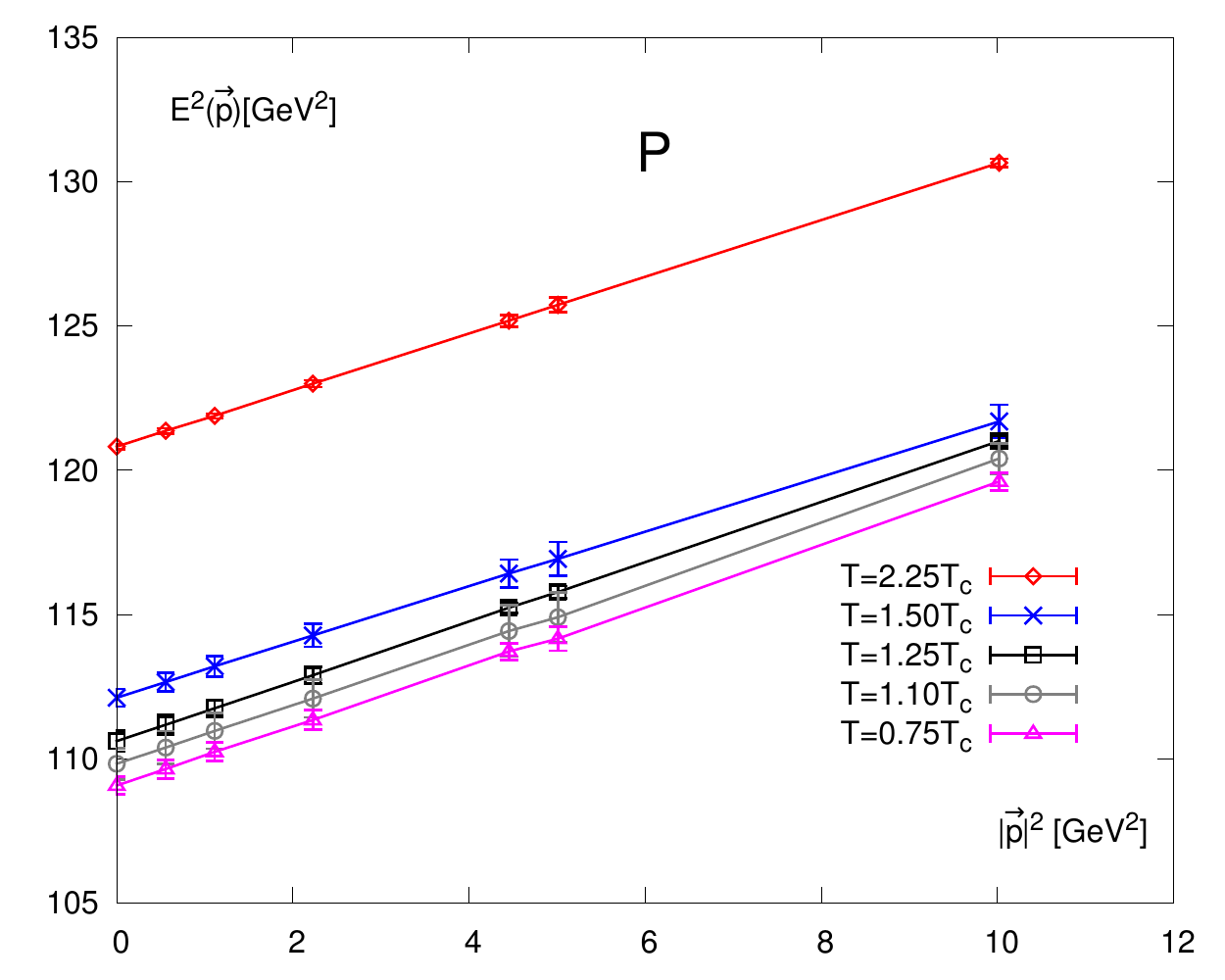}}\hfill
             {\includegraphics[width=0.475\textwidth,height=5.3cm]{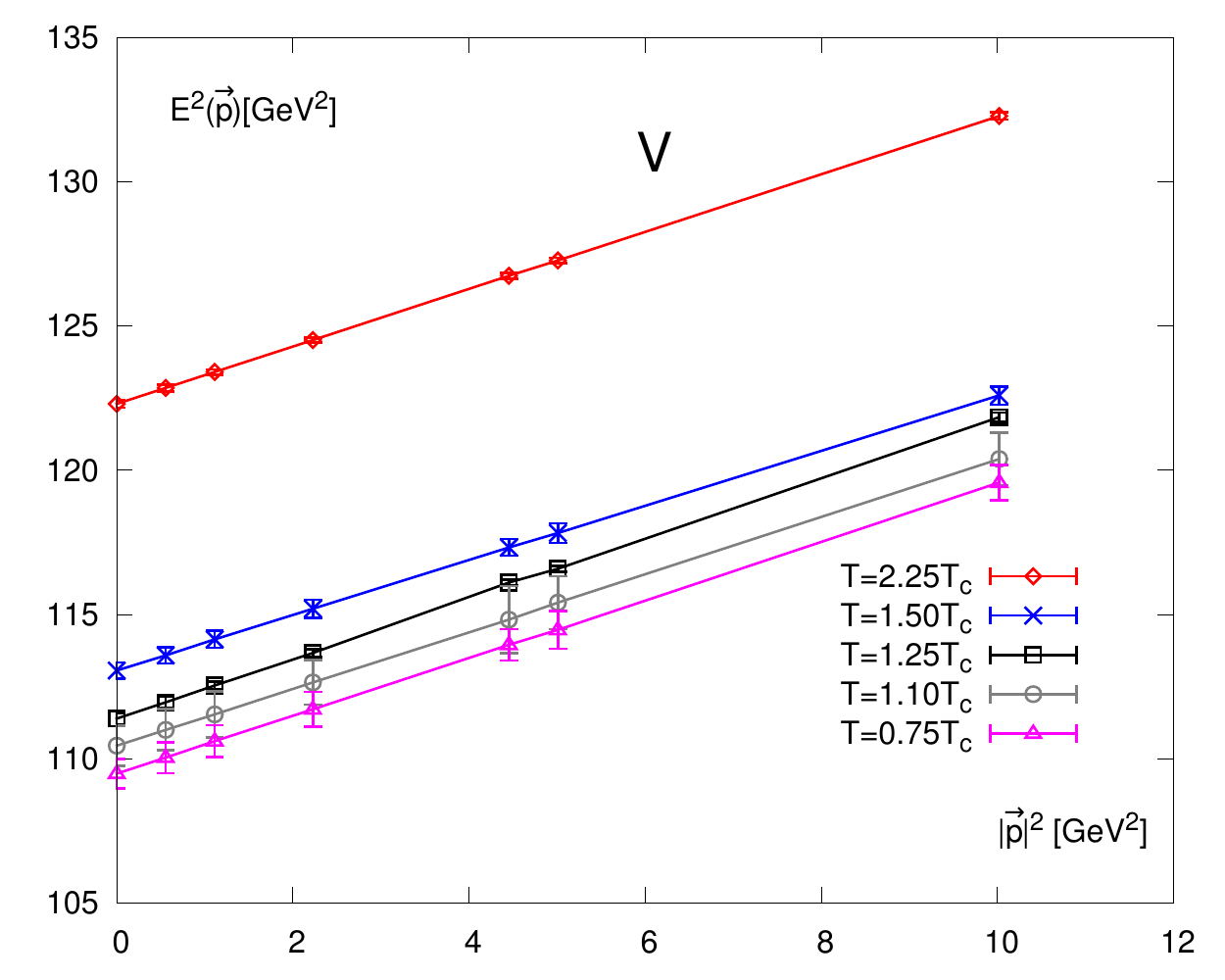}}
             {\includegraphics[width=0.475\textwidth,height=5.3cm]{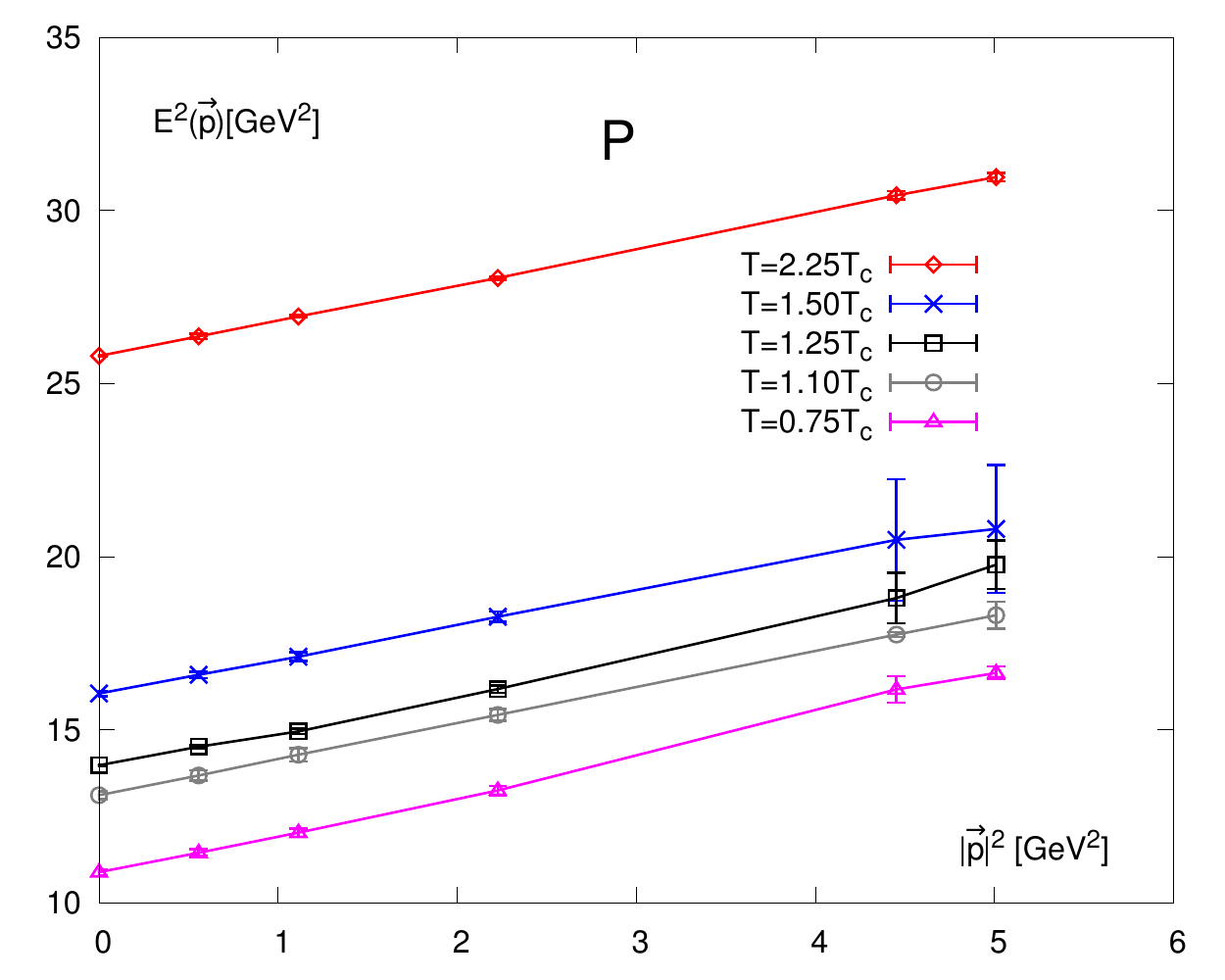}}\hfill
             {\includegraphics[width=0.475\textwidth,height=5.3cm]{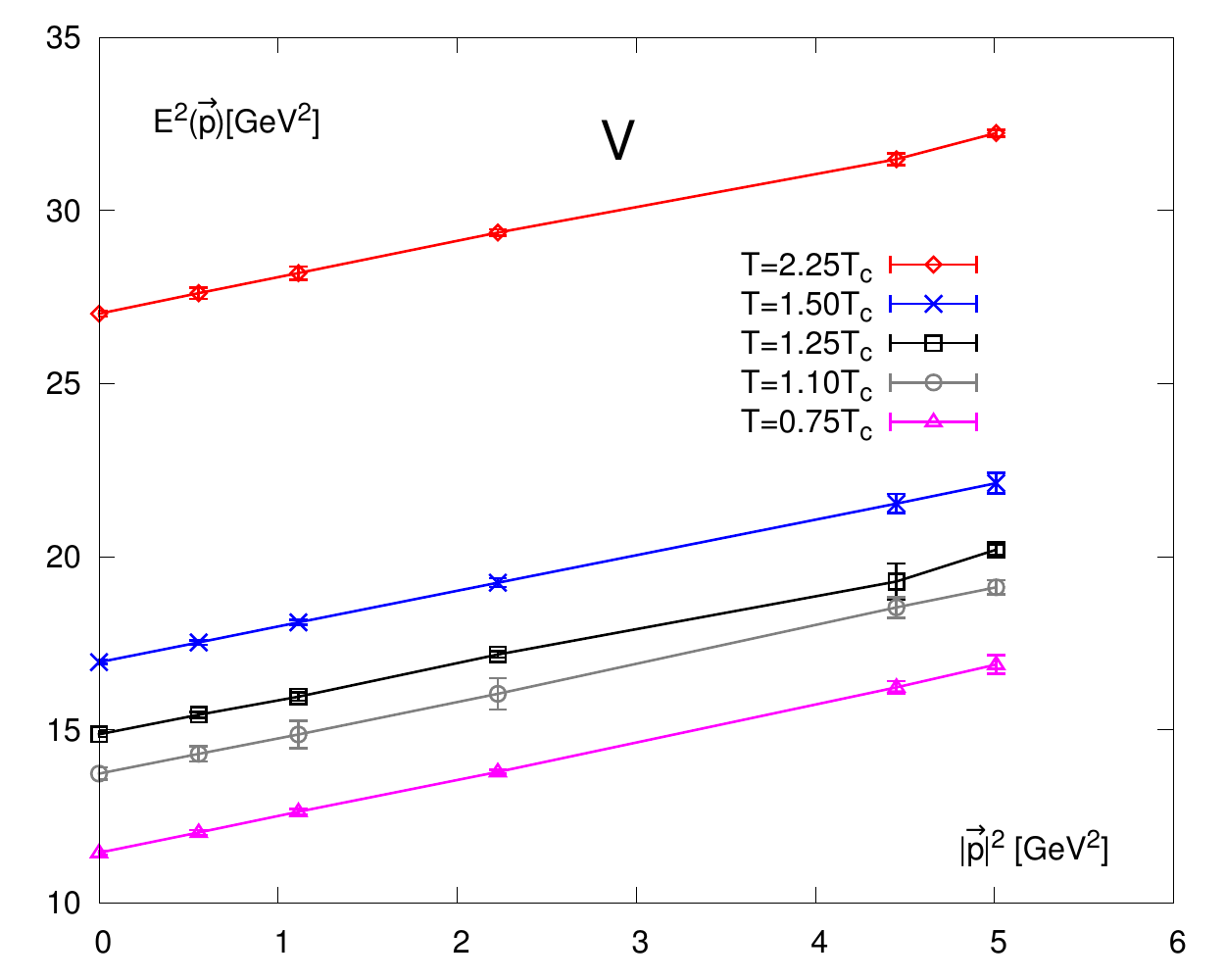}}

   \caption{Dispersion relation for botomonia (\textit{Top}) and charmonia (\textit{Bottom}) at different temperatures in the pseudo-scalar channel (\textit{Left}) and the vector channel (\textit{Right}).}
   \label{dispersion_relation}
\end{figure}

We also check the dispersion relation in the P and V channels for both charmonia and bottomonia. To do so we calculate the correlation functions at momentum $(p_x,p_y)$ ranged in $[(0,0),(1,0),…,(3,3)]\times 2\pi /(aN_{\sigma})$ and simultaneously  $\omega_{n}=0$, corresponding to $|\mathbf{p_{\perp}}|\in$[0,3.17] GeV. From the top two panels of Fig. \ref{dispersion_relation} we find that for bottomonia the dispersion relation remains linear and $A(T)\sim 1$ (see Eq. (\ref{absorb})). A similar result using non-relativistic quarks can be found in \cite{Aarts:2012ka}. As for charmonia shown on the bottom panel of Fig. \ref{dispersion_relation}, a similar situation has been observed as in the case of bottomonia, where the dispersion relation remains unmodified, which is consistent with \cite{Ikeda:2016czj,Ding:2012pt}. The reason could be that the largest momentum 3.17 GeV is still of the order of charmonium masses and less than those of bottomonia.

\section{Summary}\label{summary}

We have performed simulations on large quenched isotropic lattices to calculate the spatial Euclidean correlation functions. Applying correlated $\chi^2$-fitting we are able to extract reliable screening masses from these correlation functions. By checking the screening masses at different temperatures and zero momenta, we find that the screening masses of S-wave states for both bottomonia and charmonia increase monotonically in temperature. For bottomonia, $E_{src}(2.25T_c)/E_{src}(0.75T_c)-1$ is $5.6\%$ while for charmonia $54\%$. The screening masses of P-wave states for both bottomonia and charmonia increase non-monotonically in temperature. Our quenched calculations and 2+1 flavor HISQ calculations show the similar change tendency of the screening masses. In both our quenched simulations and 2+1 HISQ simulations, $E_{src}$ of P-wave states have a dip but they appear at different temperatures. At non-zero momenta we find that the dispersion relation in our quenched simulations seems to be not modified in the medium. The reason could be that the momentum of the quarkonium state is less than its mass at rest.

\section{Acknowledgement}\label{acknowledge}
HTS is grateful to members of HotQCD collaboration for many useful suggestions. This work is supported by the National Natural Science Foundation of China under grant numbers 11535012 and 11521064, the Deutsche Forschungsgemeinschaft (DFG) through the grant CRC-TR 211 ''Strong-interaction matter under extreme conditions’’ and the U.S. Department of Energy, Office of Science, Office of Nuclear Physics, through Contract No. DE-SC001270. The computations in this work were performed on the Aachen, Bielefeld, Juelich and Paderborn machines.

\bibliography{lattice2017}

\end{document}